\documentclass{article}
\pdfoutput=1
\usepackage[a4paper, total={6in, 8in}]{geometry}
\usepackage{graphicx}
\usepackage{multirow}
\graphicspath{ {./Images/} }
\usepackage[utf8]{inputenc}
\usepackage{multicol}
\usepackage{tabularx}
\usepackage{slashbox}
\usepackage{authblk}
\usepackage{pifont} %for the checkmark symbol

\newcommand{\fsquare}{\ding{110}}
\newcommand{\fcircle}{\ding{108}}
\newcommand{\esquare}{\ding{111}}
\newcommand{\ecircle}{\ding{109}}

\usepackage[%
  style=numeric,
  backend=biber,
  natbib=true
]{biblatex}
\title{Recent Advances in Digital Image and Video Forensics, Anti-forensics and Counter Anti-forensics}
\author{Maryam Al-Fehani}
\author{Saif Al-Kuwari}

\affil{Division of Information and Computing Technology, \\ College of Science and Engineering, \\Hamad Bin Khalifa University, \\ Qatar Foundation, Doha, Qatar.}
\date{}

\begin{document}

\maketitle

\begin{abstract}
   Image and video forensics have recently gained increasing attention due to the proliferation of manipulated images and videos, especially on social media platforms, such as Twitter and Instagram, which spread disinformation and fake news. This survey explores image and video identification and forgery detection covering both manipulated digital media and generative media. However, media forgery detection techniques are susceptible to anti-forensics; on the other hand, such anti-forensics techniques can themselves be detected. We therefore further cover both anti-forensics and counter anti-forensics techniques in image and video. Finally, we conclude this survey by highlighting some open problems in this domain.  
\end{abstract}

%\tableofcontents

%%%%%%%%%%%%%%%%%%%%%%%%%%%%%%%%%%%%%%%%%%%%%%%%%%%%%%%%%%%%%
\section{Introduction}\label{Intro}

Multimedia is commonly used as a form of communication to spread information and for entertainment purposes. Multimedia takes various forms, mainly: video, image and audio. It is highly present in news channels and social media platforms.  In fact, multimedia has become an essential element in our society, both for individuals and organizations, to connect people, provide news, and provide security and surveillance (e.g., CCTV) \cite{Sadeghi_2018}. 

However, multimedia is not always utilized with good intentions. Therefore, it is imperative to differentiate between ethically and unethically modified multimedia, such as an image that is altered for enhancement purposes rather than spreading disinformation.  Recently, the use of unethically modified multimedia has emerged as a means of changing public opinion and spreading disinformation \cite{Cozzolino_2018, Li_2018, Mccloskey_2018, Xuan_2019}.  Although false information may be spread inadvertently through misinformation, this survey aims to explore the spread of disinformation, which refers to the deliberate spread of false or fake information \cite{Hernon_1995} as indicated in Table \ref{matrix}. 

\begin{table}[!h]
\centering
\begin{tabular}{lllll}
\cline{1-3}
\multicolumn{1}{|l|} {\backslashbox{Information}{Intention}   }       & \multicolumn{1}{l|}{Malicious intention}      & \multicolumn{1}{l|}{Good Faith}     &  &  \\ \cline{1-3}\cline{1-3}
\multicolumn{1}{|l|}{Genuine Information}   & \multicolumn{1}{l|}{Obscure}        & \multicolumn{1}{l|}{Genuine}        &  &  \\ \cline{1-3}
\multicolumn{1}{|l|}{Falsified Information} & \multicolumn{1}{l|}{Disinformation} & \multicolumn{1}{l|}{Misinformation} &  &  \\ \cline{1-3}
\end{tabular}
\caption{Matrix that shows the genuineness of an image}
\label{matrix}
\end{table}

This increase in the unethical use of multimedia is mainly due to the public availability of advanced multimedia editing software, machine learning and deep learning-based tools \cite{Piva_2013, Cozzolino_2019, Bravo_2011, Dehnie_2006}. Unfortunately, multimedia has recently been used to spread disinformation online and promote fake news campaigns \cite{Cozzolino_2018, Li_2018, Mccloskey_2018, Xuan_2019,Nguyen_2019}. Even news channels have been deceived by manipulated media \cite{Jawed_2017}. %https://thewire.in/media/2017s-top-fake-news-stories-circulated-by-the-indian-media
%president gains 3 million followers in an hour, tweeted across the world from fake image

With the ease of creating manipulated multimedia of high quality, it has become difficult to distinguish between fake and authentic media \cite{Marra_2019, Nataraj_2019, Yang_2020, Mandelli_2020}. This is exacerbated in the case of images and videos, where it is no longer feasible to use human judgment to detect fake media \cite{Li_2018, Mandelli_2020,Piva_2013, Nataraj_2019, Yang_2020, Mccloskey_2018}. Malicious use of multimedia does not only influence public opinion, it can also affect the lives of targeted victims \cite{Li_2018, Meena_2019, Cozzolino_2018, Li_2018, Mccloskey_2018, Xuan_2019}. Furthermore, the difficulty in discerning the authenticity of multimedia has affected various aspects, such as legal proceedings, which rely heavily on multimedia as evidence, and military operations, which have been deceived by fake media \cite{Nguyen_2019}. 

Multimedia forensics mainly examines the integrity of media and detects any modifications. This paper focuses on multimedia forensics, focusing mainly on image and video forensics. %of three main sub-fields, namely: image forensics, video forensics and audio forensics \cite{Sadeghi_2018}. 
 Indeed, multimedia manipulation has existed since the inception of these media. An early example of image-related manipulations is shown in \cite{Meena_2019}, which demonstrates that fake images existed at least since 1840.  

\begin{figure}[ht]
\centering
\includegraphics[width=0.5\textwidth]{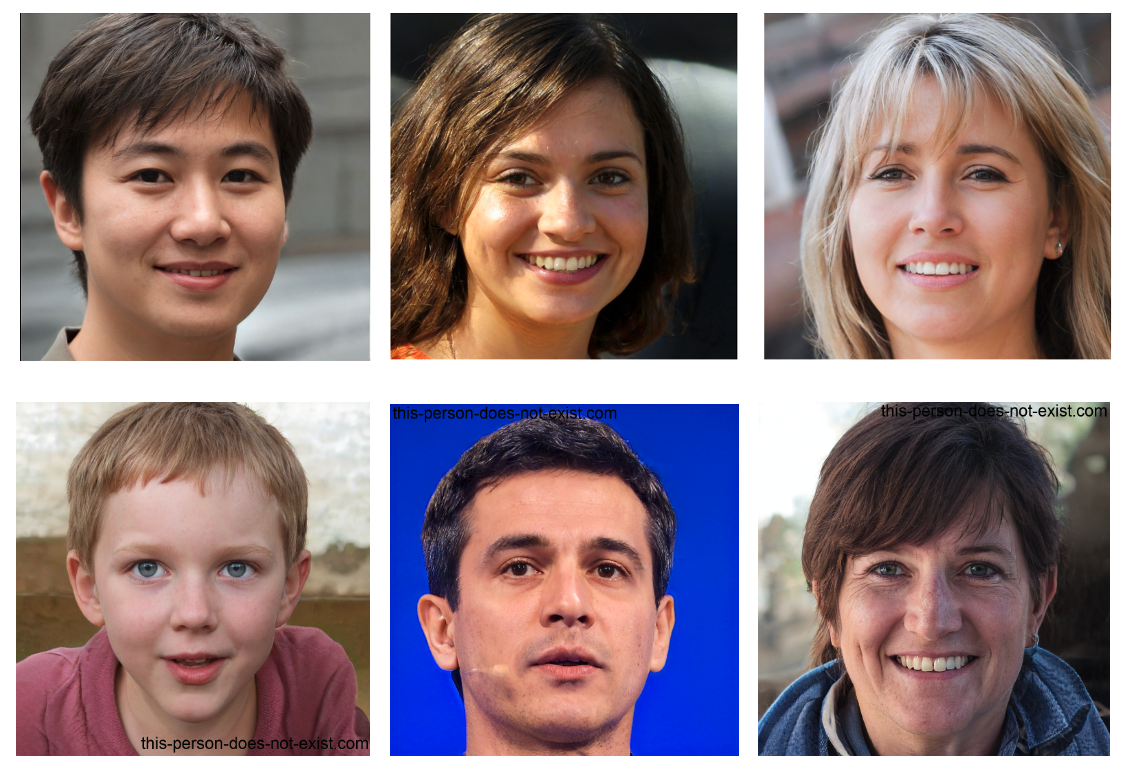}
\caption{GAN Generated Images from \cite{Wang_2019} based on StyleGAN works by \cite{Karras_2019}  \cite{Xuan_2019}.}
\label{StyleGan2}
\end{figure}

A relatively new approach to creating fake images and videos is based on a deep learning method called Generative Adversarial Networks (GANs) that can create synthesized images \cite{Goodfellow_2014}. This technology has since proliferated, with many applications. Generative images and videos have seen many advances and are now able to synthesize extremely realistic media, such as the ones shown in Figure \ref{StyleGan2}, which do not portray real people. Generative image and video algorithms have also been used to swap faces for malicious purposes in the infamous DeepFake app \cite{Nataraj_2019}. Generative images and videos are increasingly becoming more difficult to detect \cite{Antipov_2017, Gulrajani_2017, Arjovsky_2017, Karras_2017, Nguyen_2019, Salimans_2016}. Thus, the detection of generative images and videos has progressively attracted more research interest. Various surveys discussed image and video forensics as shown in table \ref{related_works}, which highlight the scope of this survey compared to the existing ones. %  \cite{Nguyen_2019}.

\begin{table}[htbp!]
\centering
\small{
\begin{tabular}{|m{2cm}|m{2cm}|m{2cm}|m{2cm}|m{2cm}|m{2cm}|}
\hline
Surveys & Source Identification & Forgery Detection & Generative Detection & Machine Learning & Counter Forensics \\
\hline
\cite{Piva_2013} & \fsquare~~\ecircle & \fsquare~~\ecircle  &  \esquare~~\ecircle & \fsquare ~~\ecircle  & \fsquare~~\ecircle \\
\hline
\cite{Van_2007} & \fsquare~~\ecircle & \fsquare~~\ecircle  &  \fsquare~~\ecircle & \esquare ~~\ecircle  & \esquare~~\ecircle \\
\hline
\cite{Luo_2007} & \fsquare~~\ecircle & \fsquare~~\ecircle  &  \esquare~~\ecircle & \fsquare ~~\ecircle  & \esquare~~\ecircle \\
\hline
\cite{Ansari_2014} & \esquare~~\ecircle & \fsquare~~\ecircle  &  \esquare~~\ecircle & \fsquare ~~\ecircle  & \esquare~~\fcircle \\
\hline
\cite{Gill_2017} & \esquare~~\ecircle & \fsquare~~\ecircle  &  \esquare~~\ecircle & \fsquare ~~\ecircle  & \esquare~~\fcircle \\
\hline
\cite{Sadeghi_2018} & \esquare~~\ecircle & \fsquare~~\ecircle  &  \esquare~~\ecircle & \fsquare ~~\ecircle  & \esquare~~\fcircle \\
\hline
\cite{Meena_2019} & \esquare~~\ecircle & \fsquare~~\ecircle  &  \esquare~~\ecircle & \fsquare ~~\ecircle  & \esquare~~\fcircle \\
\hline
\cite{Yang_2018} & \fsquare~~\ecircle & \esquare~~\ecircle  &  \fsquare~~\ecircle & \esquare ~~\ecircle  & \fsquare~~\fcircle \\
\hline
\cite{Bernacki_2020} & \fsquare~~\ecircle & \esquare~~\ecircle  &  \fsquare~~\ecircle & \fsquare ~~\ecircle  & \fsquare~~\ecircle \\
\hline
\cite{Verdoliva_2020} & \esquare~~\ecircle & \fsquare~~\fcircle  &  \fsquare~~\fcircle & \fsquare ~~\fcircle  & \fsquare~~\fcircle \\
\hline
\cite{Nowroozi_2020} & \esquare~~\ecircle & \fsquare~~\ecircle  &  \fsquare~~\ecircle & \fsquare ~~\ecircle  & \fsquare~~\ecircle \\
\hline
\cite{Barni_2018} & \fsquare~~\ecircle & \fsquare~~\ecircle  &  \esquare~~\ecircle & \fsquare ~~\ecircle  & \fsquare~~\ecircle \\
\hline
\cite{Gragnaniello_2021} & \esquare~~\ecircle & \esquare~~\ecircle  &  \fsquare~~\ecircle & \fsquare ~~\ecircle  & \esquare~~\ecircle \\
\hline
\cite{Girish_2019} & \esquare~~\ecircle & \fsquare~~\ecircle  &  \esquare~~\ecircle & \fsquare ~~\ecircle  & \esquare~~\ecircle \\
\hline
\cite{Kaur_2020} & \esquare~~\ecircle & \fsquare~~\fcircle  &  \esquare~~\ecircle & \fsquare ~~\fcircle  & \esquare~~\ecircle \\
\hline
\cite{Sharma_2019} & \esquare~~\ecircle & \esquare~~\fcircle  &  \esquare~~\ecircle & \esquare ~~\ecircle  & \esquare~~\fcircle \\
\hline
\cite{Singh_2018} & \esquare~~\ecircle & \esquare~~\fcircle  &  \esquare~~\ecircle & \esquare ~~\ecircle  & \esquare~~\fcircle \\
\hline
\cite{Kingra_2016} & \esquare~~\ecircle & \esquare~~\fcircle  &  \esquare~~\ecircle & \esquare ~~\fcircle  & \esquare~~\fcircle \\
\hline
\cite{Clark_2019} & \esquare~~\ecircle & \fsquare~~\fcircle  &  \fsquare~~\fcircle & \fsquare ~~\fcircle  & \esquare~~\ecircle \\
\hline
\cite{Shelke2_2021} & \esquare~~\fcircle & \esquare~~\fcircle  &  \esquare~~\ecircle & \esquare ~~\fcircle  & \esquare~~\fcircle \\
\hline

Ours & \fsquare~~\fcircle & \fsquare~~\fcircle  &  \fsquare~~\fcircle & \fsquare~~\fcircle  & \fsquare ~~\fcircle \\
\hline

\end{tabular}}
\caption{Comparison between other surveys and this survey, where the symbols~\fsquare~and~\fcircle~ indicate that the technique is covered in the corresponding survey for image and video, respectively. Similarly, the symbols~\esquare~and~\ecircle~indicate that the technique is not covered for image and video, respectively.}
\label{related_works}
\end{table}

\subsection{Contributions}
The contributions of this survey can be summarized as follows: 
\begin{enumerate}
    \item Provide an up-to-date review of state of the art in image and video forensics research%review of the state of the art
    \item Examine source identification and forgery detection for both images and videos
    \item Explore detection of generative images and videos
    \item Analyze the anti-forensics and counter anti-forensics methods for both images and videos
    \item Highlight the open problems and existing gaps for future research
\end{enumerate}
%%Summarize what i am doing in the survey
%%Taxonomy, state of the art
%%Future gaps

\subsection{Organization}
\begin{figure}[ht]
\centering
\includegraphics[width=1.1\textwidth]{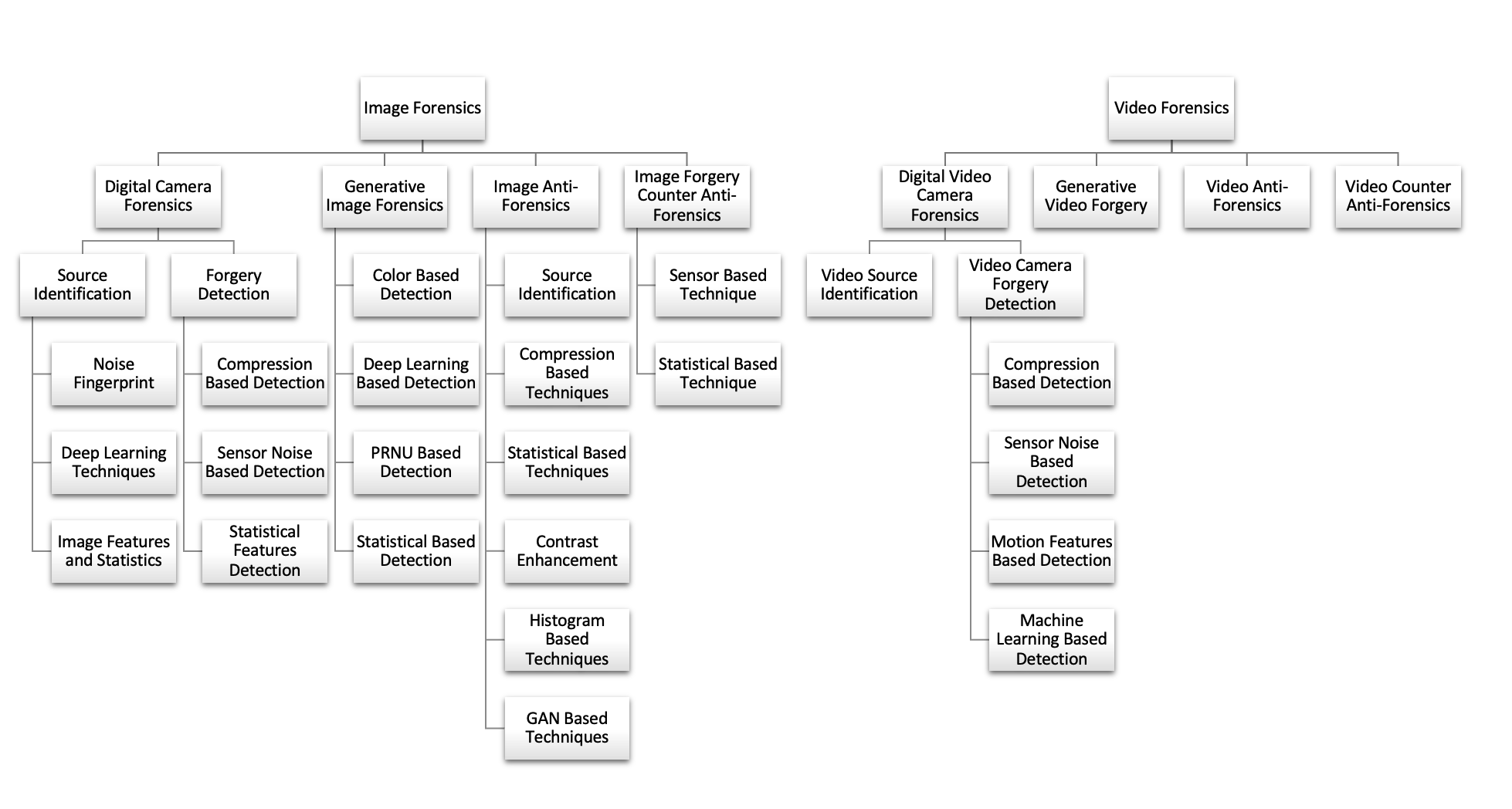}
\caption{Organization of the image and video forensics}
\label{Organization}
\end{figure}

The rest of this paper is divided into two main sections: Section \ref{imageForensics}, which focuses on image forensics, and Section \ref{videoForensics}, which discusses video forensics. In each section, we discuss detection methods (Sections \ref{camera_forensics}, \ref{image_gen}, \ref{video_detection}, and \ref{video_gen}), anti-forensics methods (Sections \ref{Anti-Forensics}, and \ref{Video_anti-forensics}) and counter anti-forensics methods (Sections \ref{Counter-Anti-Forensics}, and \ref{video_counter}). Finally, we conclude the survey in Section \ref{Problems_Conclusion} by highlighting open problems and promising future directions in this domain.  
Figure \ref{Organization} visually illustrates the organization of this survey. 
%%to add the organization chart

\section{Image and Video Datasets}\label{dataset}
%This section covers the datasets that are commonly used in both the image and video forensics fields, this is inclusive of generative forgeries, the 

Table \ref{Dataset_t} illustrates the most commonly used datasets in image and video forensics research, including generative forgery techniques.

% https://dl.acm.org/doi/abs/10.1145/3394171.3413769 -- Added
% https://arxiv.org/abs/2108.05080 -- Added
% https://openaccess.thecvf.com/content/ICCV2021/html/Kwon_KoDF_A_Large-Scale_Korean_DeepFake_Detection_Dataset_ICCV_2021_paper.html -- Added

\begin{table}[]
\centering
\begin{tabular}{|p{25mm}|p{20mm}|p{20mm}|p{25mm}|p{35mm}|}
\hline
Dataset & Media Type & Forgery Types & Dataset Size & Description \\ \hline
Celeb-df \cite{Li2_2020} (2020) Cited by $\sim 741$ references & Videos & DeepFake & 5,639 fake videos and 590 real videos & contains high quality Deepfakes of celebrities. \\ \hline
WildDeepfake \cite{Zi_2020} (2020) Cited by $\sim 164$ references & Videos & DeepFake & 7,314 fake videos & Contains synthesized face sequences derived from Deepfakes videos. \\ \hline
FakeAVCeleb \cite{Khalid_2021} (2021) Cited by $\sim 53$ references & Audio-Video & DeepFake & 500 real videos and 19,500 fake videos & Contains Deepfakes videos with fake audio. \\ \hline
KoDF \cite{Kwon2_2021} (2021) Cited by $\sim 46$ references & Audio-Video & DeepFake & 62,166 real videos and 175,776 fake videos & Contains real and synthesized Deepfakes videos that focus on Korean subjects. \\ \hline
Faceforensics++ \cite{Rossler_2019} (2019) Cited by $\sim 1435$ references& Videos & Tampering & 1000 fake videos and 1000 real videos & Publicly available dataset that was collected through YouTube and social media \\ \hline
DFDC \cite{Dolhansky_2020} (2020) Cited by $\sim 402$ references& Video & DeepFake & Over 100,000 fake videos and 20,000 real videos & Contains videos that are Deepfake, GAN-based.  \\ \hline
CoMoFoD \cite{Tralic_2013} (2013) Cited by $\sim 342$ references& Image & Copy Move Forgery & 13,000 forged images and 520 real images & contains images that have copy-move forgery, in addition to post processing and manipulations which include translation, rotation, scaling, combination and distortion \\ \hline
CASIA \cite{Dong_2013} (2013) Cited by $\sim 382$ references& Image & Splicing and Copy Move Forgery & 800 real images and 921 fake images & contains image forgeries that are for both copy move and splicing forgeries
 \\ \hline
VISION \cite{Shullani_2017} (2017) Cited by $\sim 244$ references& Video/Image & Source Device Identification & 34,427 real images and 1914 real videos & Created through 35 different devices and is available publicly\\ \hline
\end{tabular}
\label{Dataset_t}
\caption{Popular Image and Video Forgery Datasets}
\end{table}

\section{Image Forensics}\label{imageForensics}

Digital images can be broadly categorized as digital camera-generated images and Generative Adversarial Network images. Digital camera-generated images can be captured with a device and may be referred to as natural images, whereas GAN-generated images are created through a deep learning method and are not considered natural images. Generative images were popularized by Deepfakes, which were introduced in 2017 \cite{Tolosana_2020}. The use of generated faces or deep learning has since been used in multiple ways, including face synthesis, identity swap and expression manipulation. In this section, we begin by examining digital camera image forensics through source identification to identify the image's camera model and source. Then, we explore forgery detection for digital camera images, after which we examine GAN generated images forensics. Finally, we discuss anti-forensics, which is used to circumvent detection, and counter anti-forensics techniques, which are used to detect hidden forgeries.

%%%%%%%%%%%%%%%%%%%%%%%%%%%%%%%%%%%%

\subsection{Digital Camera Image Forensics} \label{camera_forensics}

Digital camera image forensics is divided into active forensics and passive forensics. The active forensics approach requires the image to include camera-dependent features, thus the digital camera must be able to insert a watermark or digital signature as part of the pre-processing cycle of image acquisition \cite{Piva_2013, Singh_2013, Arnold_2003, Lu_2003}. Active methods are considered to be less practical as these features affect the quality of the captured image and most devices are not manufactured with them. 

Accordingly, the literature shifted towards the passive methods \cite{Meena_2019}, which utilize inherent characteristics and statistics in the image that do not require any information to be inserted into the image during the capturing process. Generally, these passive methods are concerned with source identification and forgery detection. %In this section, we discuss the most popular techniques in this domain  and categorize them as illustrated in Figure \ref{Forensic_category}. 

%\vspace{5mm} %5mm vertical space

%\begin{figure}[ht]
%\centering
%\includegraphics[scale=0.37] {Images/Categories.png}
%\caption{Categorization of Passive Image Forensics}
%\label{Forensic_category}
%\end{figure}

%%Note to self, popescu_2004
\subsubsection{Source Identification}
Source identification has been an active area of research since the early 2000s. Source identification is carried out mainly through noise fingerprints (artifacts extracted from images), deep learning and image feature \& statistical techniques. Source identification techniques are summarized in \ref{Image_Source_Identification}.
%It is mainly conducted through noise fingerprints, which is an artifact extracted from images. While the deep learning method uses advanced networks to identify image sources. Source identification is also addressed through the use of statistical and image features, these statistics and artifacts can also be used to detect forgeries. 

\paragraph{Noise Fingerprint}
Sensor noise fingerprint is an artifact generated during image acquisition by physical and hardware noise \cite{Julliand_2015}. Below, we briefly describe image noise types:
\begin{itemize}
  \item Physical noise: refers to factors such as lighting, and consists of dark shot noise and photon shot noise.
  \begin{itemize}
    \item Dark shot noise: generated through an accumulation of heat-generated electrons within the camera sensor.
    \item Photon shot noise: results from how light naturally reaches the pixels in an irregular manner for two neighboring pixels.
\end{itemize}
  \item Hardware noise: refers to the noise produced due to mechanical issues within the camera, which mainly includes Fixed Pattern Noise (FPN) and Photon Response Non-Uniformity (PRNU).
   \begin{itemize}
    \item FPN: caused by sensor flaws and can be removed from the image.
    \item PRNU: caused by sensor flaws, but as it is non-linear and hence it is difficult to remove from an image.
\end{itemize}
\end{itemize}

Image noise has been identified as a method of \emph{source identification} since 2005, when Lukas \emph{et al.} \cite{Lukas_2005, Lukas_2006} were able to obtain sensor pattern (referred to as `` natural watermark" ) from images. The study utilized non-uniformity noise (PRNU) of the pixel from the image sensors as a type of image watermark. 
The researchers extracted pattern noise from images that use charge-coupled device (CCD) and complementary metal oxide semiconductor (CMOS) sensors.  Later, the authors in \cite{Filler_2008} were able to demonstrate that PRNU can identify images based on camera model and brand, since they were able to classify the fingerprints of 4,500 digital cameras with an accuracy rate of 90.8\%.  

Another work  \cite{Goljan_2009} conducted a large-scale test to detect sensor fingerprints, where the classification of over 1 million images taken from 150 models was tested. The results indicated that the error rates did not increase due to having the images taken from different cameras from the same model. This means that the fingerprint extraction methodologies utilized are accurate and effective. However, they found that image quality affected the fingerprint extraction method, as the lower the image quality, the less accurate the extraction is. Similarly, in \cite{Liu_2010}, the authors improved the noise fingerprint and decreased the false rejection rate due to extracting PRNU. 

More recently, the authors in \cite{Gupta_2018} focused on improving the performance of sensor pattern noise by suppressing peaks and low-frequency defects to decrease the false acceptance rate.  However, a study \cite{Meij_2018} of the use of PRNU to detect sources of images that are shared on social media platforms concluded that it is not viable to determine the source camera identity shared on social media through sensor fingerprints. 
 
\paragraph{Deep Learning Techniques}
The use of deep learning for source identification is a relatively new approach and appears to have been introduced in \cite{Baroffio_2016} and \cite{Tuama_2016} as they used CNNs (Convolutional Neural Networks) to identify cameras and camera models. A similar CNN-based approach \cite{Bondi_2016} was proposed to identify both unique camera devices as well as camera models. These methods were not able to identify the unique camera accurately, unlike the PRNU method, but they were able to identify the camera model of the images with an accuracy of 94.1\%. Later, the authors in \cite{Yao_2018} were able to further improve camera model identification by restructuring the CNN model based on extracted image features. 
%This paper utilized a multi-classifier in place of a binary classifier which allowed it to simultaneously classify a large range of images with a high accuracy (approaching 100\%); However, the classifier wasn't able to accurately detect specific camera models from certain manufacturers.

Another approach utilized CNN to detect source camera model \cite{Kuzin_2018} by training an enhanced DenseNet architecture to process low-level image features that are robust against gamma, resize, contrast and resize transformations. This approach was able to identify the camera model of images with an accuracy of 98\%. More recent research explored camera models, device identification and camera brand identification for mobile devices \cite{Ding_2019}, which resulted in 84.3\% accuracy of device identification. The method attempts to take advantage of past research that utilizes both hand-crafted, data-driven techniques and a domain knowledge-driven method.
Similarly, the authors in \cite{Freire_2019} focused on identifying the source mobile camera to enable the detection of both the device manufacturer and the unique camera used to capture an image. The authors used a CNN that uses noise patterns to detect and identify the device model and camera. They were able to detect the device manufacturer with an accuracy of 98.1\% and the unique camera with 91.1\% accuracy. Furthermore, another approach was introduced in \cite{Bayar_2018} to address the open set camera model identification problem, which aims to classify unknown cameras by proposing the use of a CNN with a classifier that relies on confidence score mapping and two binary classifiers trained to identify unknown camera models. The approach was able to resolve the open set problem with an accuracy of 97.74\% accuracy.

Another direction is camera model identification with post-processing modification, such as editing or forgery. The authors in \cite{Rafi_2019} proposed a deep learning algorithm that utilizes augmentation and CNN built on the DenseNet architecture to identify the source camera model even in the case that the image has been through post-processing. The method achieved 96.66\% accuracy of model identification with the placement of post-processing changes. 
Similarly, the authors in \cite{Yu_2020} utilized a pre-trained CNN and trained it on a dataset containing compressed images with known manipulations; the CNN model was able to identify the source of manipulated images.  Another CNN-based method was proposed in \cite{Mayer2_2018} to validate whether images were captured from the same camera model. The model achieved an accuracy of 84-95.8\% depending on whether the image model source was known or unknown. %Recently, a multi-view multi-scale deep learning supervised method was proposed in \cite{Chen_2021} to identify generalizable features and detect manipulated images.

\paragraph{Image Features and Statistics}

Source identification is commonly achieved by utilizing various statistics and image feature-based detection methods, including JPEG quantization, chromatic aberration, color filter array (CFA interpolation noise, also called demosaicing) and other intrinsic image features. 

The authors in \cite{Farid_2006} created a JPEG quantization database that contains quantization values based on numerous camera models and the values of images that had undergone editing on commonly used editing tools. It was found that some cameras have similar quantization values and could be grouped together, indicating that this method may still be used to narrow down the image source. 

On the other hand, a different method \cite{Van2_2007}, which utilizes chromatic aberration, to identify mobile phone camera source has been used with an accuracy between 86-87\%. However, these experiments were conducted on a relatively small number of images and camera models and appear to be susceptible to counter-forensics techniques we will discuss in section \ref{Anti-Forensics}.

Another method utilizes demosaicing to detect whether an image source is a camera or computer-generated \cite{Gallagher_2008}. The authors proposed the detection of traces of demosaicing as an indicator of an image captured from a digital camera. However, demosaicing authentication method relies on the hypothesis that computer generated images do not exhibit any demosaicing patterns and could be misled by a color filter on the generated image.  Additionally, a co-occurrence matrix based SVM classifier was proposed in \cite{Marra_2015} to identify the image's camera model. The proposed algorithm is based on extracting the image features based on the co-occurrence matrix and was able to achieve accuracy up to 99.44\%.

\begin{table}[!h]
\centering
%\scalebox{0.7}{
\begin{tabular}{|p{2.5cm}|p{3.5cm}|p{3.5cm}|p{2.5cm}|} 
\hline
Technique & Detection Approach & Pros & Cons    \\  \hline
Noise Fingerprint   & PRNU \cite{Lukas_2005, Lukas_2006, Filler_2008, Goljan_2009, Liu_2010, Gupta_2018, Meij_2018}  & Accurate, admissible in legal proceedings &  Susceptible to generated images using noise fingerprints    \\ \hline
Deep learning   techniques       & Customized   CNN \cite{Baroffio_2016, Tuama_2016, Bondi_2016, Yao_2018, Freire_2019, Bayar_2018, Yu_2020, Mayer2_2018, Chen_2021}, DenseNet \cite{Kuzin_2018, Rafi_2019}, ResNet \cite{Ding_2019} &          Automated method, ability to detect image sources &  Susceptible to AF, may not be admissible in legal proceedings sources      \\ \hline 
Statistical and Image Features & interpolation \cite{Gallagher_2008}, Co-occurence Matrix \cite{Marra_2015}, Image Quantization \cite{Farid_2006}, Chromatic Aberration \cite{Van2_2007} & Able to detect image source with high accuracy &  Susceptible to AF   \\ \hline                              
\end{tabular}
%}
\caption{Overview of Digital Camera Source Identification}
\label{Image_Source_Identification}
\end{table}
%\vspace{5mm} %5mm vertical space

\subsubsection{Forgery Detection}

%\subsubsection{Image Tampering Techniques}\label{Tampering}

Forgery detection in images examines two main tampering techniques \cite{Gill_2017, Luo_2007}: dependent and independent tampering. Dependent tampering includes image splicing, copy move and anomaly detection, while independent tampering mainly focuses on image retouching and light inconsistency \cite{Piva_2013, Meena_2019, Sadeghi_2018}.  Both independent and dependent tampering may be used separately or in combination, and may result in malicious forgery that alters the meaning of the original image \cite{Luo_2007}.

Image forgeries are mainly performed through image editing tools. When performed to spread disinformation, it alters the meaning of the image. Tampering techniques can be performed individually on an image or they can be combined, making their detection even more challenging. Below we briefly describe the most common types of tampering techniques \cite{Bourouis_2020}, 

\begin{itemize}
  \item Copy-Move: copy at least one region of an image and place it onto another region(s) of the same image.
  \item Image Splicing: copy at least one region of one image onto another image.
  \item Image Resampling: utilize geometric transformations (scaling and rotation) to manipulate the image.
  \item Image Retouching: commonly used after other tampering techniques are utilized on an image, to enhance the quality of the manipulated image and hide signs of tampering.    
\end{itemize}

Image forgeries can be detected through compression-based, statistical features based and machine learning based methods, the forgery detection techniques are shown in Table \ref{Image_Forgery}.

\begin{table}[]
\centering
\begin{tabular}{|p{20mm}|p{35mm}|p{35mm}|p{25mm}|p{25mm}|}
\hline
Forgery   Technique & Description & Detection   Technique & Pros & Cons \\ \hline
Copy-Move & Copy at   least one region of an image and place it onto another region(s) of the same image. & Block-based \cite{Bravo_2011, }, Key-based \cite{Huang_2008, Ardizzone_2015, Wang_2017, Li_2014, Wenchang_2016, Yang_2018, Silva_2015}, Hybrid of block and key based \cite{Zhao_2015}, Dense Field \cite{Cozzolino_2015}, Lens and Chromatic Aberration  \cite{Yerushalmy_2011, Mayer_2018}, Statistical Methods \cite{Hsu_2008, Huang_2017} & Accurately detects copy move & It is limited to copy-move tampering, it is susceptible to AF \\ \hline
Image   Splicing & Copy at least one region of one image onto another image. & Optical and motion \cite{Rao_2014, Bahrami_2015}, Machine Learning \cite{AlAhmadi_2013, ElAlfy_2015, Shen_2017}, Statistical Methods \cite{Zhao_2014, Park_2016, Singh_2021, Pun_2016}  & Can   accurately detect image splicing & It is limited to only to image splicing, it is susceptible to AF \\ \hline
Anomaly   Detection & Used to detect novel tampering techniques and the use of multiple manipulations in an image. & Image dimple \cite{ Agarwal_2017}, Robust Hashing \cite{Tanaka_2021}, Statistical Methods \cite{ Fan_2015, Johnson_2006, Li2_2018, Tanaka_2021}, Deep Learning \cite{ Mayer_2020, Wu_2019}  & Can detect image forgeries, can be used to detect a variety of forgeries & It is susceptible to AF \\ \hline
Image   Retouching & Commonly used after other tampering techniques are utilized on an image, to enhance the quality of the manipulated image and hide signs of tampering. & Statistical Methods \cite{ Lin_2013, Padin_2017, Cao_2014, Popescu_2005, Li_2010, Lien_2010, Qian_2012, Stamm_2008, Stamm_2009, Stamm2_2010}, Deep learning \cite{ Bayar2_2017, Cao_2011, Ding_2018}, Machine Learning \cite{ Cao_2009, Gaikwad_2021}, Deep Learning \cite{ Jingyu_2018, Zhuo_2018, Boroumand_2018} Histogram based methods \cite{ Guo_2018} & Can   accurately detect image forgeries & It is susceptible to AF \\ \hline
Light   Inconsistencies & Uses light inconsistencies to detect certain forgeries, mainly image splicing. & Statistical and Geometric Analysis \cite{ Johnson_2005, Saboia_2011, Wu_2011, Kee_2013, Kee_2010, YanLi_2014}, Deep learning \cite{ Pomari_2018}, Illuminant Maps \cite{ Carvalho_2013, Carvalho_2016} & Can accurately detect image forgeries & Limited to a few forgery techniques, it is susceptible to AF \\ \hline
\end{tabular}
\label{Image_Forgery}
\caption{Overview of Digital Camera Forgery Detection}
\end{table}

\paragraph{Compression Based Detection.}
%utilizes the use of compression as a method of forgery detection

The compression-based methods rely on compression features as a means to determine the occurrence of forgeries. Some authors \cite{Lin_2009,Thing_2012} proposed to detect JPEG tampered images by analyzing double quantization effect. Similarly, the authors in \cite{Liu_2014} utilized an algorithm that uses a special descriptor based on JPEG block artificial grid and noise estimation to detect copy-move and splicing forgery. In addition, another work \cite{Bianchi_2011} proposed to detect image forgery based on JPEG compression non-alignment through a statistical method that automatically detects forgeries in a DCT block. 

Recently, an approach was proposed \cite{Singh_2021} that can detect image splicing amongst other forgeries by analyzing images to determine the use of different JPEG compression quality factors. This method demonstrated that it is possible to detect image splicing and is resilient to other techniques such as rescaling and copy-move manipulations. Additionally, the authors in \cite{Milani_2014} proposed an algorithm based on the analysis of DCT coefficients that can detect multiple JPEG compressions to determine forgeries. Similarly, the work in \cite{Thai_2019} utilized DCT coefficient analysis to detect double compression.

Another work \cite{Galvan_2014} proposed an algorithm that utilizes successive quantization to identify double compression. Additionally, the authors in \cite{Bianchi2_2011} proposed a method to estimate primary quantization to detect forgeries. Furthermore, the work in \cite{Bianchi_2012} improves on this method by proposing an algorithm that automatically computes the likelihood of double compression based on image blocks to further localize tampering. Additionally, the work in \cite{Li2_2015} introduced a novel method that utilizes a quantization noise-based approach to detect JPEG decompression.

\paragraph{Sensor Noise Based Detection.}
%this needs to be covered

The sensor noise-based methods rely on image sensor noise to determine forgeries. The work in \cite{Lukas_2006} introduced a method to detect forgeries by determining the presence of sensor pattern noise and identifying forged regions that were formed by the manipulated noise pattern. The authors in \cite{Chierchia_2011} utilized PRNU to detect small tampering through the use of segmentation-based analysis. Additionally, the work in \cite{Chierchia_2014} utilized sensor pattern noise through Markov random field and convex optimization techniques to detect forgeries. Another work \cite{Chierchia2_2014} proposed a technique that uses PRNU in addition to a spatially adaptive filtering technique to detect forgeries that include small forgeries.

Another work \cite{Cozzolino_2014} proposed an algorithm that combines sensor noise, machine learning and block matching along with implementing a decision fusion strategy to detect forgeries. Another work in \cite{Korus_2016} also utilized decision fusion and tampering probability maps of various sizes to detect small and large forgeries. The authors in \cite{Gardella_2021} introduced a stochastic model to identify image forgeries by detecting local noise anomalies. Another work \cite{Lin_2020} utilized PRNU homogeneity and multi-scale image segmentation to detect and localize forgeries. %This method was tested on public realistic tampering image dataset and was found to outperform state of art detectors. 

\paragraph{Statistical Features Based Detection.}
%statistical and pixel, geometric and physics, inconsistencies (light)
The statistical features-based detection methods rely on statistical features, image pixels, geometry and physics to detect forgeries.  In this section, we discuss the following methods: Copy-Move Detection, Image Splicing Detection, Anomaly Detection, Image Retouching Detection, Light Inconsistencies based Detection, and Machine Learning Based Detection. 

\begin{enumerate}

\item \textit{Copy-Move Detection:}
Copy-move tampering is one of the most widely used tampering techniques. The authors in \cite{Bravo_2011} studied the use of block-based detection, in which the image is divided into blocks that are analyzed to detect duplications. However, further analysis illustrates that geometric manipulations such as rotation of a copied object can hinder the block-based approach. To address this concern, the authors created a one-dimensional descriptor of the block that does not get altered based on geometric manipulations. Another technique, called keypoint, was introduced in \cite{Huang_2008} in conjunction with the SIFT-based framework to detect copy-move forgeries. The keypoint method is used to extract image patches that are then matched to detect forgeries. Additionally, another study  \cite{Zhao_2015} combined the two techniques to identify the manipulated segment of the image whilst addressing geometric manipulations. Similar approaches are also proposed in \cite{Zhao_2015, Zandi_2016}.
Another approach \cite{Ardizzone_2015} was able to detect copy-move by using matching triangles of key-points through vertex descriptors. However, this method does not perform well with complex images. While this method provides a novel approach to matching triangles and uses less computational complexity than the block-based approach, it has a high rate of false matches. Another method \cite{Wang_2017} utilizing key-point extracts uniform key-points was proposed that uses SURF detector while considering the probability density gradient. It introduces a superpixel content-based adaptive feature points detector, robust EMs-based key-point feature and more efficient based key-point matching. However, this method exhibits high computational complexity. 
Similarly, another method \cite{Li_2014} segments the image prior to applying the key-point based detection technique. Moreover, a new algorithm was proposed \cite{Wenchang_2016} based on particle swarm optimization, to automatically adapt the parameter values and improve CMF SIFT based detection. While the authors in \cite{Yang_2018} propose a novel key-point distribution strategy that addresses the lack of keypoints in image. This method is able to detect forgeries even when other post-processing methods (e.g. scaling, rotation and blurring) are applied to the image. 
Relying on multi-scale analysis, another algorithm \cite{Silva_2015} changes the color space of the image, then detects and clusters interest points. The method relies of Speeded-Up Robust features and nearest neighbor distance ratio and is able to detect forgeries in a robust manner as it is able to overcome tampering techniques which include image rotation and resizing. However, the authors indicate that the detection method may not find enough key-points and may therefore not be able to make correct detection. Similarly, another approach \cite{Silva_2015} proposes the use of pre-processing, key-point matching as well as multi-scale analysis and voting based on detection maps. The method is able to detect copy-move even with the presence of scaling and rotation with an accuracy of up to 96.19\%. 
Another algorithm \cite{Huang_2017} delves into detecting copy-move in JPEG compressed images through the use of FFT (Fast Fourier Transform), SVD (Singular Value Decomposition) and PCA (Principal Component Analysis) and was able to achieve an accuracy above 97\%.  Furthermore, an algorithm that relies on dense field techniques and the PatchMatch algorithm was introduced in \cite{Cozzolino_2015} to rapidly generate the approximate nearest neighbor field for the whole image. 
Another approach proposed in \cite{Yerushalmy_2011} uses lens aberration as a detection method; Lens aberration results from lens and CCD sensors. The work indicates the potential for copy-paste forgery detection as the proposed algorithm was able to detect tampered areas of the image. A recent work \cite{Mayer_2018}  created a statistical model based on chromatic aberration to detect forgeries with higher accuracy compared to \cite{Hsu_2008} which uses a fusion of statistical analysis to detect forgeries.
However, it is important to note that this high detection rate was only attainable through one of the camera models examined. 

%\subsubsection{Image Splicing Detection}
\item \textit{Image Splicing Detection}: 
Numerous techniques were proposed for image splicing forgery detection, including: optical blur, motion blur, Markov features, multi-scale noise estimate and co-occurrence matrices \cite{Rao_2014, Bahrami_2015, ElAlfy_2015, Zhao_2014, Pun_2016, Park_2016, Shen_2017}. The optical and motion blur methods detect image splicing through blur inconsistencies in an image, where the manipulated segment of an image appears irregular in comparison to the remaining part of the image. 
As such a method was proposed \cite{Rao_2014} to analyze spliced forgeries added to images that already have motion blur. The authors propose an automated method to detect splicing that relies on space-variant settings in an image with motion blur. Similarly, other work \cite{Bahrami_2015} also studied optical and motion blur and proposed a framework that relies on human decision to infer the blur type. Although this method shows potential, it is based on human judgment, which can limit its utility. 
%
%\begin{figure}[ht]
%\centering
%\includegraphics{Images/Figure6.png}
%\caption{A spliced image with motion blur \cite{Rao_2014}.}
%\label{MotionBlur}
%\end{figure}
%
A novel technique \cite{AlAhmadi_2013} to detect image splicing which utilizes DCT and LBP (Local Binary Pattern) as well as a SVM classifier was able to detect splicing forgeries with an accuracy of 97\%. Another method based on the Markov feature was introduced in \cite{ElAlfy_2015}, which relies on attaining spatial and discrete cosine transform (DCT) domains from an image through the Markov process in addition to the use of a SVM classifier and PCA to reduce computational complexity. The method was able to achieve an accuracy of 98.2\%. 
Additionally, another approach \cite{Zhao_2014} proposed the use of a 2-D Noncausal Markov model, which considers the current node in the image and its nearest four neighbors; the method was able to achieve 90.1\% detection rate of spliced images. Another approach was proposed by Pun \emph{et al.} \cite{Pun_2016}, which examines the use of noise estimation to detect image splicing through the use of a Simple Linear Iterative Clustering (SLIC) algorithm. The method relies on finding inconsistencies of the noise in the image and is able to identify multiple splice manipulations.  
Another approach to image splicing detection was proposed in \cite{Park_2016} which utilizes local and global statistics and co-occurrence matrices. The proposed model utilizes 144 splicing detection features and results in a detection accuracy of 95\%. Another method \cite{Shen_2017}, which relies on co-occurrence matrices, utilizes textural features based on grey level co-occurrence matrices as well as an SVM classifier and has indicated a detection rate of 98\%. 
Another approach was proposed in \cite{Singh_2021} which can detect image splicing amongst other forgeries and analyze images to find different JPEG compression quality factors. This method demonstrates that it is able to detect image splicing and is resilient to other techniques such as rescaling and copy-move manipulations.

%\subsubsection{Anomaly Detection}
\item \textit{Anomaly Detection:}
Another class of detection techniques utilizes general detection methods to detect anomalies (i.e. forgeries). The authors in \cite{Agarwal_2017} were able to identify an image feature called dimple which appears as a single darker or brighter pixel due to quantized DCT coefficients to intensity-space. The dimple occurs every 8x8 block and is introduced in 67\% of commercial cameras. The authors found that manipulations can distort the dimple feature or erase it. The JPEG dimple is resilient to some forms of post-processing such as double compression, gamma correction, additive noise and scaling.
Another novel approach was introduced in \cite{Fan_2015}, which utilizes patch likelihood and GMM (Gaussian Mixture Model) to detect if a block exhibits any tampering or post-processing traces. Furthermore, a deep learning method called forensic similarity graphs \cite{Mayer_2020} was used to analyze images for localized tampering, which may detect various image forgery techniques. Another approach was introduced recently in \cite{Tanaka_2021}, which utilizes robust hashing techniques to create a database of reference hashes that can be used to identify forgeries in both camera-generated images and generative images. 
Additionally, the authors in \cite{Johnson_2006} propose the use of chromatic aberration (CA). The authors created a model for estimating aberration and conducted experiments on both synthetic and genuine images (compressed and uncompressed), divergence from the model is indicative of image doctoring. The technique can be used to detect both copy-move and splicing forgeries. Another approach \cite{Wu_2019} which proposes the use of a self-supervised learning network is able to detect manipulated images and localize complex forgeries including 385 manipulation types. Another work \cite{Li2_2018} proposed the use of a multi-class classification scheme to identify various image operations through the use of statistical changes that occur on images when any operations are conducted. 

%\subsubsection{Image Retouching Detection}
\item \textit{Image Retouching Detection:}
Image retouching includes various techniques, this paper mainly focuses on image resampling, scaling, contrast enhancement and sharpening. Resampling forgery detection was established after statistical analysis on tampered images found indicators of tampering  \cite{Popescu_2005}. This was further expanded by \cite{Li_2010} by analyzing the periodicity of the resampling technique on a given image. Another approach \cite{Lien_2010} utilizes a pre-calculated resampling weighting table to detect image forgery with an accuracy up to 95\%.  
Another novel approach \cite{Qian_2012} uses rotation-tolerant method to detect resampling through the use of resampling history detection algorithm, with an accuracy of 97.2\%. Another approach investigated the use of CNN to detect resampled images \cite{Bayar2_2017} that are JPEG compressed to identify re-compressed images, with an accuracy up to 97.88\%. However, the authors found that when the image is downscaled the accuracy is decreased to 84.02\%. Furthermore, a detector was proposed in \cite{Padin_2017}to recognize resampling through the use of the law of eigenvalues in noise space.
On the other hand, other methods to detect contrast enhancement in images through the use a blind forensics method to detect the use of both local and global contrast enhancement as well as analyzing the histogram entries to determine what aspect of the image had been altered \cite{Stamm_2008, Stamm_2009, Stamm2_2010}.  
Additionally, another proposed method detects cut-and-paste forgeries through the use of contrast enhancement \cite{Lin_2013}, as images that undergo cut-and-paste attacks require contrast enhancement to hide light inconsistencies of the pasted images. The method achieved good results. However, it has been tested on uncompressed images and performed poorly with JPEG compression. Additionally, another work \cite{Cao_2014} conducted a similar investigation into contrast enhancement for both uncompressed images and previously compressed images and was able to accurately detect both local and global enhancements. However, the method requires further enhancement to include JPEG compressed images. 
In addition to contrast enhancement, other work \cite{Cao_2009} was conducted on image sharpening, as image sharpening may be utilized as a last step of image forgery to hide forgery traces.  The first investigation into image sharpening detection  utilized both histogram gradient metrics and ringing artifacts analysis to detect whether an image was sharpened with the assistance of a linear classifier, the method reached detection accuracy of up to 93.9\%. However, they appear to be susceptible for anti-forensics techniques. 
Accordingly, another approach \cite{Cao_2011} proposed to detect image sharpening, with a focus on the Unsharp Masking (USM) Adobe Photoshop feature. They were able to detect unique overshoot artifacts associated with the utilization of USM and were able to accurately detect utilization of sharpening on small sized images. Additionally, another work \cite{Ding_2018} proposed a novel method to detect weak sharpening strength. The approach uses edge perpendicular ternary coding (EPTC) method, which was able to detect the use of both weak and strong sharpening strength with an accuracy of at least 95\%. 
Additionally, another approach  \cite{Ding_2018} utilized CNN to detect USM with detection of approximately 98\% and above. The authors compared their detection method with EPTC \cite{Jingyu_2018} and found that the performance of the proposed CNN method is at least 10.09\% higher than EPTC. 
A novel approach was introduced to detect a new problem in the field, fake colorized images, where an adversary may colorize an image to hide objects. In \cite{Guo_2018}, the authors aimed to detect these manipulated images through FCID-HIST (Histogram based Fake Colorized Image Detection) and FCID-FE (Feature Encoding based Fake Colorized Image Detection). Other works also address this problem through deep learning \cite{Zhuo_2018} and machine learning \cite{Gaikwad_2021}. Another approach proposes \cite{Boroumand_2018} the use of a deep learning method, CNN, to detect various image retouching techniques. This method was able to detect four types of retouching which include blurring, sharpening, denoising and histogram related changes such as contrast enhancement.

%%%%%%%

%\subsubsection{Light Inconsistencies}
\item \textit{Light Inconsistencies based Detection:}
Light inconsistency was introduced as a forgery detection method in \cite{Johnson_2005} and was motivated by the work in  \cite{Nillius_2001} which explored the use of light source direction in images. The authors in \cite{Johnson_2005} utilized light direction to manually uncover manipulated images, which is mainly used for image splicing techniques, as images that were spliced together will show objects with different lighting conditions. The work in \cite{Saboia_2011} built on \cite{Johnson_2007} and studied the effect of light reflection in an individual’s eyes in images as a means of detecting image manipulations with 94\% detection accuracy. 
Additionally, another approach \cite{Wu_2011} examines the use of illuminant color inconsistencies by dividing the image into blocks and comparing the estimated illuminant color of each block to uncover manipulated blocks. Furthermore, another method \cite{Kee_2013} analyzes the shadows in images caused by light sources to determine whether image tampering was conducted. They develop a geometric method to analyze whether the shadows are consistent with one main source of light . 
Additionally, a technique was proposed \cite{Kee_2010} to analyze shadows in 3D lighting environments. This method was able to determine lighting inconsistencies in a 3D model, that considers a person’s head and body and the shadows they create in a 3D environment . Similarly, another proposed method \cite{YanLi_2014} utilizes illuminant chromaticity in RGB and its consistency in different objects in an image to detect image tampering. This method was able to detect copy-and-paste as well as scaling tampering techniques, however the method is limited by various elements such as the object’s reflectivity. 
Furthermore, the use of illuminant maps to detect splicing forgeries was explored in \cite{Carvalho_2013, Carvalho_2016} with a focus on individuals faces. It was able to utilize lighting inconsistencies, along with color, texture and shape cues and machine learning to detect forgeries, with an accuracy reaching up to 94\% for cut-and-paste face forgery. Furthermore, another approach \cite{Pomari_2018} employed Deep Learning, a SVM classifier and illuminant maps to detect image splicing, reaching an accuracy of 96\% in light inconsistency forgery detection. The authors also indicate the potential of utilizing transfer learning to further improve the results of their approach. 

\item \textit{Machine Learning Based Detection:}
%machine and deep learning methods
%
The machine learning based detection techniques utilize both machine and deep learning methods to detect forgeries. A novel technique proposed in \cite{AlAhmadi_2013} detects image splicing which utilizes DCT and LBP (Local Binary Pattern) and the use of SVM classifier to detect splicing forgeries with an accuracy of 97\%. Another method based on the Markov feature was introduced in \cite{ElAlfy_2015}, which relies on attaining spatial and discrete cosine transform (DCT) domains from an image through the Markov process in addition to the use of an SVM classifier and PCA to reduce computational complexity. The method was able to achieve an accuracy of 98.2\%. Another method \cite{Shen_2017} relies on co-occurrence matrices, utilizes textural features based on grey level co-occurrence matrices as well as an SVM classifier and has a detection rate of 98\% for image splicing forgeries. 
Another approach was introduced in \cite{Fan_2015} that utilizes patch likelihood and GMM (Gaussian Mixture Model) to detect if a block exhibits any tampering or post-processing traces. More recently, a deep learning method called forensic similarity graphs \cite{Mayer_2020} was used to analyze images for localized tampering, which may detect various image forgery techniques. Another approach \cite{Wu_2019} which proposes the use of a self-supervised learning network is able to detect manipulated images and localize complex forgeries including 385 manipulation types. Another approach in \cite{Bayar2_2017} investigated the use of CNN to detect resampled images that are JPEG compressed to identify re-compressed images, with an accuracy up to 97.88\%. However, the authors found that when the image is downscaled the accuracy is decreased to 84.02\%.
Another approach  \cite{Ding_2018} utilized CNN to detect image sharpening through USM with a detection of approximately 98\% and above. The authors compared their detection method with EPTC \cite{Jingyu_2018} and found that the performance of the proposed CNN method is at least 10.09\% higher than EPTC. Additionally, other works utilized deep learning \cite{Zhuo_2018} and machine learning \cite{Gaikwad_2021} to detect colorized images that may be used to hide objects. Another approach proposes \cite{Boroumand_2018} the use of a deep learning method, CNN, to detect various image retouching techniques. This method was able to detect four types of retouching which include blurring, sharpening, denoising and histogram related changes such as contrast enhancement.
Furthermore, the use of illuminant maps to detect splicing forgeries was explored in \cite{Carvalho_2013, Carvalho_2016} with a focus on individuals' faces. It was able to utilize lighting inconsistencies, along with color, texture and shape cues and machine learning to detect forgeries, with an accuracy reaching up to 94\% for cut and paste face forgery. Furthermore, another approach \cite{Pomari_2018} employed Deep Learning, a SVM classifier and illuminant maps to detect image splicing, reaching an accuracy of 96\% in light inconsistency forgery detection. The authors also indicate the potential of utilizing transfer learning to further improve the results of their approach. Additionally, a multi-view multi-scale deep learning supervised method was proposed in \cite{Chen_2021} to identify generalizable features and detect manipulated images.

\end{enumerate}

\subsection{Generative Image Forensics}\label{image_gen}

Deep generative learning was not widely used for image generation until 2014, when Goodfellow \emph{et al.} \cite{Goodfellow_2014} proposed a novel approach, the Generative Adversarial Network (GAN). The approach utilizes two networks, called generator and discriminator. These networks  are trained on a large dataset, where the generator attempts to synthesize fake images to look photorealistic, while the discriminator determines whether the generated image can be categorized as real or fake. This process repeats until the image produced by the generator is classified as realistic by the discriminator. 
Accordingly, various generative model developments have been made to create high-quality images (as shown in Figure \ref{StyleGan2}) including the introduction of VAE (Variational Auto-Encoder) as an image generative model \cite{Gregor_2015, Razavi_2019}. Other applications and enhancements include face swapping and face aging \cite{Antipov_2017, Gulrajani_2017, Arjovsky_2017, Karras_2017, Nguyen_2019, Salimans_2016, ElGammal_2017}. The techniques used in GAN image generation have evolved significantly and are now comparable to legitimate images in terms of quality and realism. Thus GAN forensics has been at the forefront of the recent literature. 

\paragraph{Color Based Detection}
Detection techniques for GAN generated images vary as some authors attempt to detect GANs through statistics and image features while others rely on deep learning techniques. The authors in \cite{Mccloskey_2018} studied the color cues in both real and GAN generated images. They analyzed a GAN architecture to find the use of color in the image and were able to exploit that analysis for the detection of the GAN image. However, the success rate of this approach was not high.

\paragraph{Deep Learning Based Detection}

Another approach \cite{Nataraj_2019} uses CNN along with pixel co-occurrence matrices that are computed through color channels of images to detect GAN-based images. Their proposed method was tested on two datasets and was able to achieve a detection accuracy of 99\%. A similar approach is also proposed in \cite{Goebel_2021}. To further enhance this method and outperform the work in \cite{Nataraj_2019}, the authors in \cite{Barni_2020} use CNN and both cross-band and spatial co-occurrence analysis to detect GANs. Another approach \cite{Tariq_2018} used a custom CNN, ShallowNet, to detect both human-created and GAN-generated fake faces, it had 94-99\% AUROC (area under the receiver operating characteristic) when detecting GAN-generated images.

Another approach \cite{Yang_2021} to derive a GAN fingerprint with the use of both qualitative analysis of GAN fingerprints and a multi-task generative network. The proposed network was able to detect the use of deep generative images in both close-world and open-world scenarios, with a respective accuracy of 99.99\% and 78.22\% .  Additionally, another approach \cite{Li_2018} focuses on using two main methodologies to detect GAN-generated images, namely intrusive and non-intrusive techniques.  The intrusive detector reported more accurate detection rate.

To address the detection of unseen GAN manipulations, the work in \cite{Cozzolino_2019} introduced transfer learning based method based on CNN detector, while \cite{Xuan_2019} proposed the addition of smoothing or noise to remove low-level pixel statistics in the training data, so classifier can focus on the more intrinsic aspects of the image. However, this method suffers from low accuracy, but the true negative rate is improved by 10\% when pre-processing is used.

Another approach \cite{Yang2_2021} proposes the use of an active forgery detection method to detect Deepfake images. This method proposes the addition of a deep learning based watermark to images before they are posted to social media to enable the analysis and detection of Deepfakes generated based on the image. Another work \cite{Jain_2020} introduces the use of a deep learning algorithm that separately detects generated images and retouched images and it shows the reason of the decision. In the case of a generated image, it is able to identify the GAN architecture used.

\paragraph{PRNU Based Detection}

The authors in \cite{Marra_2019} investigated whether GAN generated images have distinguishable fingerprints similar to PRNU. They conducted an experiment initially on two GANs, cycle and progressive GANs, and extracted the image’s fingerprints in a manner that is similar to PRNU. The approach was tested on various GAN architectures illustrating that GAN fingerprints can be extracted and utilized to obtain distinctive prints.  Similarly, in \cite{Yu_2019}, the authors studied the potential of GAN fingerprints highlighting that the extracted fingerprints may be susceptible to anti-forensics techniques.  

%called ForensicTransfer, which can detect unseen manipulations. This detector utilizes autoencoder-based representation learning and latent space to preserve all image representation. The latent space is further divided into two parts, one activated by fake samples while the other by real samples. Unknown samples are determined by how close they are to the real or manipulated samples. The experiments focus on manipulations from Computer Graphics (CG) or deep learning methods. The paper creates its own dataset of 30,000 images utilizing progressive GAN, cycle-GAN, style-GAN, glow and StarGAN, with different resolutions for various manipulations. They also create inpainted images, as well a CG-based dataset. This paper created an advanced method for GAN and CG recognition covering a wide range of manipulations. 

%On the other hand, 

 \paragraph{Statistical Based Detection}

Similar work has been conducted to find general properties in deep generated facial images \cite{Chai_2020}. Similarly, an algorithm \cite{Bonettini_2021} that utilizes Benford's law to distinguish generative images from camera images was proposed. This method mainly relies on analyzing DCT coefficients to make its determination. Although it had some potential, it was not able to detect all cases and it may be learned by a generative model to avoid detection. Another work \cite{Guarnera_2020} proposes the use of Expectation Maximization (EM) to extract facial features to identify GAN-generated faces. This algorithm was able to detect GAN-generated human faces with a maximum accuracy of 99.31\%.

A novel multi-scale spatio-frequency wavelet-based detection method \cite{Wolter_2021} was proposed to identify deep generative images with an accuracy that reaches up to 98.6\%. Another approach\cite{Guo_2021} was proposed to detect GAN-generated faces by identifying irregular pupil shapes through an automated method. Similarly, another work \cite{Guo2_2021} proposes the use of eye inconsistencies in GAN-generated images to detect generative images. This work utilizes a deep network to find these inconsistencies with an accuracy that reaches up to 97\%.

 Another approach \cite{Li_2020} was introduced to distinguish deep network generated images in which digital camera images and deep network generated images were compared and analyzed. The analysis shows a difference in the chrominance components of the image and was able to identify generated images even when the generative model is unknown.  Additionally, another approach \cite{Hu_2021} demonstrates that deep generated images can be detected through the use of a corneal specular inconsistency between the eyes of the generated face.

%%%%%%%%%%%%%%%%%%%%% Table will be added

%%%%%%%%%%%%%%%%%%%%% Reorganizing and reviewing the tables and the final part before placing them

\subsection{Image Anti-Forensics} \label{Anti-Forensics}

While image forensics aims to detect anomalies and extract evidence from images, anti-forensics aims to distort the findings of forensic detectors and conceal image forgeries. In \cite{Marra2_2018, Gragnaniello_2018}, adversarial attacks were analyzed and it was shown that CNNs are indeed susceptible to attacks.  In this section, we discuss the most prominent anti-forensics techniques.

Image anti-forensics (AF) explores methods that can deter image forgery detection by concealing traces of manipulation techniques, such as compression, noise fingerprints, resampling, sharpening and contrast enhancement. The source identification AF techniques can be used to hide the identity of the camera’s source and provide anonymity. This technique is often used to protect the identity of photographers, journalists and human rights advocates.

\paragraph{Source Identification}
One approach \cite{Sengupta_2017} aimed to provide source anonymization through the use of a median filter, potentially making it difficult for detectors to discern the source of the image. Likewise, another work \cite{Andrews_2020} proposes using a generative network to anonymize images. Similarly, other works \cite{Chen_2018, Guera_2017} use GAN and statistical changes to alter images to deceive CNN detectors. Another approach \cite{Chen_2020} focuses on camera trace erasing through a novel hybrid loss method conducted without disturbing the image content. A similar approach \cite{Zeng_2015} proposes to remove the camera fingerprint (PRNU) to hide an image's source. Another method \cite{Javier_2017} focuses on smartphone image source identification anti-forensics and is based on erasing and manipulating an image's PRNU to falsely associate it with another device.

\paragraph{Compression Based Techniques}
Other approaches introduce techniques that hide image compression history \cite{Stamm_2010, Cao_2015, Li_2015}. A novel method was proposed in \cite{Barni_2015} and was able to disguise multiple compressions in an image by altering the first order statistics to make the image appear to have gone through a single compression.  Furthermore, the use of an optimization algorithm was proposed \cite{Priya_2018} to eliminate signs of JPEG compression. This method enables the image manipulator to produce compressed images that are more difficult to detect from a forensics viewpoint.  Moreover,  a framework that can estimate the image’s coefficients before compression was proposed \cite{Stamm_2011}; it was then used to hide traces of compression. 
Another work \cite{Fan_2013} further explores enhancing JPEG anti-forensics by improving the visual quality of the compressed image making it more robust to detection techniques. Furthermore, the use of a dictionary-based approach was proposed in \cite{Afshin_2016} in which the patches of input image are used to eliminate JPEG compression artifacts. This is used  to disguise forgeries and hide JPEG compression. Additionally, a novel approach \cite{Luo_2018} was introduced for to deter JPEG compression forensics with the use of GAN, this method was able to deter detectors and improve the image’s quality, a similar approach was also introduced in \cite{Wu_2020}. Another approach \cite{Singh_2017} proposed the use of denoising algorithms to improve JPEG AF, but with high computational cost.

\paragraph{Statistical Based Techniques}
Researchers also examined anti-forensics methods to deter the detection of contrast enhancement (CE) in manipulated images. The methods introduced in \cite{Cao_2010, Cao2_2014} deters the forensic detectors in \cite{Stamm_2008, Stamm_2009, Stamm2_2010} which are able to determine whether an image was enhanced through the use of first order statistics, namely peak-gap artifacts in the histogram. The authors in \cite{Cao_2010, Cao2_2014} propose the use of local random dithering to hide image artifacts and add noise to the enhanced image without lowering the quality of the image. Another work \cite{Kirchner_2008} focused on re-sampling anti-forensics that utilizes undetectable image operations to deter the forensic detector introduced in \cite{Popescu_2005}.

\paragraph{Contrast Enhancement}
Additionally, the work in \cite{Ravi_2016} proposes an ACE (Anti-Forensics Contrast Enhancement) method to deter the enhanced detector that relies on second order statistics \cite{DeRosa_2015}. The ACE method has been shown to degrade both the first and second order statistics methods used to detect CE. Similarly, another work \cite{Sharma_2020} proposes a method that hides traces of both CE and median filtering, through circumventing spatial domain forensics detectors with the use of an optimization method. Furthermore,  another approach \cite{Zou_2021} explores the use of GAN in combination with histogram-based loss and pixel-wise loss to hide traces of CE images, this results in CE manipulated images that are difficult to detect. 

\paragraph{Histogram Based Techniques}
Another popular approach is histogram based anti-forensics techniques, which essentially conceals any histogram altering manipulations from forensics detectors. An approach \cite{Barni_2012} explores a universal technique to hide histogram traces, as it obscures manipulations from first order statistics based forensics detectors through altering the histogram to a level of normalcy that makes it undetectable. The authors in \cite{Alfaro_2013} further enhance the proposed universal technique with a focus on hiding traces of JPEG compression. Another approach \cite{Wu2_2019} proposes the use of WGAN-GP (Wasserstein generative adversarial network with gradient penalty) to add multiple manipulations on an image. This method was tested against image forensics detectors and was found to be resilient against detection.  

\paragraph{GAN Based Techniques}

Another approach \cite{Cozzolino_2021} uses GAN to generate images and then inject traces of a camera model. The method was able to deceive both image source and GAN image detectors to identify the generated image as a real camera generated image. Generally, hiding traces of multiple manipulations in an image is challenging, and was addressed in several works \cite{Wu_2021, Xie_2021, Wang_2021, Huang_2020, Carlini_2020} through the use of GANs to hide such traces with minimal effect on the image's quality. 

%A similar approach is proposed in \cite{Xie_2021}. Similarly, an approach that focuses on developing GAN generated images to become less detectable is introduced. This method was able to deter detection of various forensics detectors \cite{Wang_2021}. Similarly, another method was proposed to reduce the detectability of deepfakes through reduction of GAN related artifacts during the creation of an image. This method was able to deter several state of the art detectors \cite{Huang_2020}. Additionally, another algorithm was proposed and was able to significantly reduce detection of generated images by making simple changes such as flipping the lowest bit of each pixel \cite{Carlini_2020}.

\subsection{Image Counter Anti-Forensics}\label{Counter-Anti-Forensics}

To resist anti-forensics techniques, counter anti-forensics (CAF) techniques were investigated. It was shown in \cite{Marra2_2018, Gragnaniello_2018} that methods, such as CNNs, are susceptible to attacks. Accordingly, the research community responded with novel techniques to counter anti-forensics and create more robust forensic techniques \cite{Pandit_2014, Singh_2016, Gul_2017, Böhme_2013}. In this section, we discuss the most prominent counter anti-forensics techniques.

%Several works address camera identification namely, noise camera identification. 

\paragraph{Sensor Based Technique}
In  \cite{Goljan_2010}, the authors explore how to detect implanting of sensor fingerprints onto a forged image without fingerprints and were able to show that the placement of a fake sensor fingerprint leaves traces that can be used to determine if an image was manipulated. 

\paragraph{Statistical Based}
Another algorithm \cite{DeRosa_2015} tackles counter anti-forensics for contrast enhancement through second-order statistics analysis to determine  the use of contrast enhancement. The algorithm utilizes co-occurrence matrix and can determine whether contrast enhancement was used on images that utilize AF techniques such as \cite{Barni_2015}.

Another work \cite{Singh_2019} proposes the use of co-occurrence matrices as well as second-order statistics. This method is able to detect median filtering and contrast enhancement. To further improve contrast enhancement CAF, a CNN based algorithm is presented in \cite{Sun_2018} that utilizes a  GLCM (gray-level co-occurrence matrix), this method is further enhanced to detect CE in JPEG compressed images in \cite{Shan_2019}. 

Several works address the noise dithering technique proposed by \cite{Stamm_2011}, including \cite{Li2_2012} which proposes a method that uncovers JPEG dithering. The authors were able to re-compress the manipulated image into different compression factors and measure the noisiness of the image and the total variation which resulted in a 97\% accuracy detection rate. Additionally, another work \cite{Valenzise_2013} proposes a method to counter the dithering technique proposed which leverages the added noise pattern and analyzing the DCT coefficients not subject to the dithering technique, this methods results in an accuracy of 93\%.

Similarly, a wavelet-based compression CAF algorithm \cite{Wang_2014} was proposed that uses DWT (Discrete Wavelet Transform) histogram and were able to detect AF wavelet compressed images. Moreover, other works  \cite{Jiang_2013, Zeng_2018} utilize noise estimation and game theory to address the noise dithering technique with detection accuracy above 99\%. 

Additionally, a CAF method that uses statistical correlations and analysis was introduced in \cite{Li_2012} to detect manipulated images subject to noise dithering, a similar approach was also proposed by \cite{Valenzise_2013}. Furthermore, another proposed method  \cite{Fahmy_2016} uses spatial frequency phase variations to analyze and compare image blocks to determine if an image had been manipulated. The various blocks are compared to uncover inconsistencies as a means of recognizing dithering attempts.  

Another approach \cite{Bhardwaj_2018} uses two detectors to determine the use of blocking artifacts through the correlation of blocks and DCT coefficients.  Further research was conducted on double JPEG compression CAF by \cite{Barni_2016}, as they counter the universal method proposed by  \cite{Barni_2015} that hides traces of multiple compressions. Accordingly, another work proposes an adversary aware data driven detector that is designed to counter the mentioned universal technique \cite{Barni_2016}. 

Another line of research in CAF addressed image resampling. A semi non-intrusive Blackbox method \cite{Cao_2012} was proposed to identify resampling operations. The authors analyze images to determine whether they utilize traditional resampling operations or anti-forensic operations and determine the type of anti-forensics used and they propose an identification procedure for such resampling operations. Additionally, to address the anti-forensics approach proposed by \cite{Kirchner_2008} which removes periodic artifacts with irregular sampling, the work in \cite{Peng_2015} shows that resampling creates new artifacts on the manipulated image due to the addition of periodic sampling and the interpolation steps. The authors use these artifacts to detect the anti-forensics resampling techniques, they further enhance their method in \cite{Peng_2017} to detect both resampling and anti-forensics resampling in images.

Furthermore, a general counter anti-forensics technique that relies on an auto-regressive model was proposed \cite{Zeng_2016} to detect various anti-forensics techniques. This method is unique as it has multiple purposes and can address JPEG anti-forensics, median filtering anti-forensics, resampling anti-forensics and histogram based anti-forensics. Another work \cite{Fontani_2014} utilizes data fusion to counter anti-forensics as it uses various image forensics and CAF tools to detect AF attempts. The paper focused on image splicing and has shown to be an effective strategy to find AF manipulated images. In Table \ref{ImageAntiForensics}, the forensic techniques and the image anti-forensics and counter anti-forensics methods are illustrated.

\begin{table}[]
\begin{tabular}{|p{30mm}|p{45mm}|p{30mm}|p{30mm}|}
%\hline
\hline
Forensics   Technique & Description & Anti-Forensics & Counter   Anti-Forensics \\ \hline
Source   Identification (PRNU) & Detects an   image’s source device or camera model & Median Filter \cite{ Sengupta_2017}, GAN \cite{ Andrews_2020, Chen_2018, Guera_2017}, Statistical method \cite{ Chen_2020, Zeng_2015, Javier_2017} & Statistical Methods \cite{ Goljan_2010}\\ \hline
Compression   History & Detects the   compression history of an image as an indicator of forgery. & Statistical method \cite{ Barni_2015, Stamm_2011, Fan_2013, Afshin_2016, Singh_2017}, Optimization Algorithm \cite{ Priya_2018, }, GANs \cite{ Luo_2018, Wu_2020}  & Statistical Methods \cite{ Li2_2012, Valenzise_2013, Wang_2014, Jiang_2013, Zeng_2018, Li_2012, Valenzise_2013} \\ \hline
Contrast   Enhancement & Detects   whether the image has undergone contrast enhancement as part of image   manipulation. & Statistical method \cite{ Cao_2010, Cao2_2014, Ravi_2016}, Optimization Algorithm \cite{ Sharma_2020}, GANs \cite{ Zou_2021} & Statistical Methods \cite{ DeRosa_2015, Barni_2015, Singh_2019},  Deep Learning \cite{ Sun_2018, Shan_2019, Bhardwaj_2018, Barni_2016, Barni_2015} \\ \hline
Histogram   Based Techniques & Detects   changes to an image’s histogram artifacts, which is used as an indicator for multiple   manipulations. & Statistical method \cite{ Barni_2012, Alfaro_2013}, GAN \cite{ Wu2_2019} & - \\ \hline
Resampling & Detects when an image has undergone the resampling process. & Statistical Method \cite{ Kirchner_2008}, GAN \cite{ Cozzolino_2021} & Statistical Method \cite{ Cao_2012, Peng_2015, Peng_2017} \\ \hline

Multiple   Manipulations & Detects the use of multiple manipulations. & GAN \cite{Wu_2021, Xie_2021, Wang_2021, Huang_2020, Carlini_2020} & Statistical Methods \cite{ Fahmy_2016, Zeng_2016, Fontani_2014} \\ \hline

\end{tabular}
\caption{Overview of Anti-Forensics and Counter Anti-Forensics Techniques}
\label{ImageAntiForensics}
\end{table}

%%%%%%%%%%%%%%%%%%%%%%%%%%%%%%%%%%%%%%%%
%%%%%%%%%%%%%%%%%%%%%%%%%%%%%%%%%%%%%%%%
%%%%%%%%%%%%%%%%%%%%%%%%%%%%%%%%%%%%%%%%
\section{Video Forensics} \label{videoForensics}
Generally, Digital video can be captured through digital cameras or artificially generated via Generative Adversarial Networks. Digital videos are a series of images that are captured from a video camera along with audio and other elements \cite{Shelke_2021}. Although videos were considered to be trusted media that could not be edited easily, that is not the case in recent years. The availability and accessibility of advanced video editing software made editing and generating video almost effortless. Like images, generative videos were popularized by Deepfakes and are used to forge videos or synthesize realistic fake videos. In this section, we first explore digital camera video forensics through source identification. Then we examine forgery detection for digital camera video and GAN generated video forensics. Finally, we examine video anti-forensics and video counter anti-forensics techniques.

\subsection{Digital Video Camera Forensics}\label{video_detection}
%%%%% generally introduce active and passive techniques and their main types
Video camera forensics mainly involves source identification and forgery detection. Source identification is used to identify the source of a digital camera video. Forgery detection techniques can either be active or passive. Active detection techniques require the original video to include embedded features, and include digital signature, intelligent technique and watermark \cite{Habeeb_2019, Shelke_2021}. Passive detection techniques detect forgeries based on features that are found within any video and include spatial, temporal and spatio-temporal techniques.

%%%
\subsubsection{Source Identification}
Identifying video camera sources is imperative as it is used to conduct criminal investigations.  The work in \cite{Iuliani_2019} used camera sensor pattern noise to discern the video source camera through sensor fingerprints of extracted video frames. A similar approach was proposed in an earlier work \cite{Van_2009}. Another work \cite{Taspinar_2019} proposes the use of PRNU in conjunction with the ratio of alignment to identify the source of a video. Similarly, another work utilized PRNU and Spatial Domain Averaged (SDA) for source attribution \cite{Taspinar_2020}. Additionally, the authors in \cite{Althamneh2_2016} propose the use of green channel PRNU (G-PRNU) to identify the source of a video, showing that this approach provides better results than the traditional PRNU method. 

In addition, the authors in \cite{Altinisik_2020} address the problem posed by stabilized videos as they deter the PRNU identification method due to the post-processing operations, such as cropping and warping. Thus, the authors integrate the spatial variants into their technique to improve the attribution accuracy of source identification up to 19\% above other attribution methods proposed at the time. Similarly, the work in \cite{Bellavia_2019} identifies the source of stabilized videos through capturing PRNU in various frames of a video and conducting a comparative analysis. Another work addresses the stabilized video identification in \cite{Mandelli_2019}. Another approach proposed in \cite{Chen_2013, Chen_2014} focused on authenticating wirelessly streamed videos which are vulnerable to blocking and blurring due to their nature. The authors propose the use of wireless channel signatures and selective frame analysis to identify the source of the video.

Furthermore, another work \cite{Kirchner_2019} utilizes deep learning with sensor pattern noise for source attribution. Similar work was also conducted in \cite{Mandelli2_2020, Verdoliva_2019}. Another approach is proposed in \cite{Altinisik_2021} to improve source identification attribution of compressed videos through the use of a novel approach that combines PRNU and block based weighting. Additionally, the authors in \cite{Altinisik_2022} utilized file metadata to identify the video’s source with an accuracy of 91\%.  Another work proposes to address the source identification problem in \cite{Martin_2022} through the use of three identification methods including voting, pattern correlation and PCE vectors. The voting method showed the most potential, as it identifies the source from each video frame and utilize a majority vote or occurrence of the identified source to identify camera source.

Additionally, the authors in \cite{Mandelli3_2020} propose the use of frequency domain parameters to identify the video source, the method resulted in an accuracy reaching up to 97\%. The authors in \cite{Khelifi_2017} propose the use of perceptual hashing to both identify and authenticate videos. It can authenticate videos even with the presence of manipulations. Other works aim to identify the social media video source through the use of various methods, the authors in \cite{Maiano_2021} propose a novel solution that utilizes both machine and deep learning techniques to identify the social media platform which videos have been shared through. Similarly, deep learning was utilized to identify the source social media platform in \cite{Amerini_2021}.

\subsubsection{Forgery Detection}
%%Categorized as mainly:
%Compression based
%Noise Artifacts
%Motion Features
%Statistical Features
%Machine Learning based

%papers without category 
%\cite{Chen_2017, Chittapur_2019, Damiano_2018, Nguyen2_2019}

Digital video forgery is available for both digital camera and generated videos, which may be manipulated in a traditional manner as illustrated in Table \ref{videoForgery}.  In Table \ref{videoForgery}, digital camera video forgery detection techniques are illustrated. Thus, we briefly describe the most common types of tampering techniques used for digital video camera forgeries below \cite{Habeeb_2019, Shelke_2021}:

\textbf{Spatial based forgery:} %intraframe
These forgeries manipulate the content of certain video frames and they include the below techniques:
\begin{itemize}
  \item Copy-move: This technique allows the attacker to add  from a video scene. 
  \item Inpainting: The removal of an object entails the use of inpainting which repairs the image frame after removing an object. This technique is also used to replicate objects in the same frame of a video or other frames of the video.
  \item Splicing: This technique entails the creation of a new video frame through pasting a part of a frame to another video frame.
  \item Upscale Crop: This method crops the edges of a video frame to hide an object.
\end{itemize}

\textbf{Temporal Forgery:} %interframe
These forgeries manipulate the order of the video frames and they include the below techniques:
\begin{itemize}
  \item Frame Deletion and Insertion: Certain frames are either inserted into the video or removed in a manner that changes the meaning of the video.
  \item Frame Duplication: In which video frames are duplicated in a different segment(s) of the video.
  \item Frame Shuffling: This technique alters the order of frames from the original order, altering the meaning of the video.
\end{itemize}

\textbf{Temporal-Spatio Forgery:}
Includes a combination of the spatial and temporal based forgeries.

The video camera forgery detection field encompasses various detection techniques to identify forgeries, these mainly include compression based, sensor noise based, motion features based, statistical and machine learning based methods. 

\begin{table}[]
\begin{tabular}{|p{25mm}|p{45mm}|p{30mm}|p{25mm}|p{25mm}|}
\hline
Forgery  Detection Technique & Description & Forgery Techniques & Pros & Cons  \\ \hline
Compression Based & Compression features are used to determine the authenticity of videos and forgery attacks. & Spatial Forgery \cite{Ravi_2014, Su_2015} and Temporal forgery \cite{Fadl_2018, Fadl2_2018, Gironi_2014, Jiang_2013} & It can be used to detect spatial and temporal forgeries. & It has not been used for spatio-temporal detection.   \\ \hline
Sensor Noise Based & Sensor noise is used for source identification and forgery detection & Spatial Forgery \cite{Chetty_2010, Hsu2_2008, Hyun_2013, Kobayashi_2010, Pandey_2014, Singh_2017, Singh2_2017} & It can be used to detect spatial forgeries. & It has not been used for temporal and spatio-temporal detection and It has not been used for generalized forgery detection.   \\ \hline
Motion Features Based & Utilizes time dependent features to detect forgeries. & Spatial Forgery \cite{Bidokhti_2015, Karthikeyan_2020, Kingra_2017, Tan_2015}, Temporal forgery \cite{Su_2009, Chao_2012, Dong_2012, Feng_2014, Singh3_2017} and Spatio-Temporal \cite{Bestagini_2013, Bozkurt_2017, Jia_2018} & It can be used to detect various forgery methods. & It has not been used for generalized forgery detection.  \\ \hline
Statistical Based & Utilizes video frame attributes and pixels to detect forgeries. & Spatial Forgery \cite{Aloraini_2019,Aloraini_2020,Chitradevi_2014,Mathai_2016,Su_2017, Su_2019, Tralic_2014}, Temporal forgery \cite{Lin_2012, Jia_2015, Ulutas_2018, Xu_2016, Yin_2014} and Spatio-Temporal \cite{Althamneh_2016} & It can be used to detect various forgery methods. & It may be susceptible to AF \\ \hline
Machine Learning Based & Detects   when an image has undergone the resampling process. & Spatial Forgery \cite{Dvino_2017, Johnston_2020, Saddique_2019, Zampoglou_2019}, Temporal forgery \cite{Jaiswal_2013, Shanableh_2013} & It can be used to detect forgeries in a video frame and for temporal forgery detection. & It has not been used for spatio-temporal detection. \\ \hline
\end{tabular}
\caption{Overview of Video Detection Techniques}
\ref{videoForgery}
\end{table}

\paragraph{Compression Based Detection}
%utilizes the use of compression as a method of forgery detection

Various works rely on the use of compression as a means to determine the authenticity of videos by determining whether they had undergone forgery attacks. The work in \cite{Fadl_2018} utilizes standard deviation of residual frames to determine whether the video had undergone a temporal forgery and is able to localize the forgery. Similarly, the proposed algorithm in \cite{Fadl2_2018} focused on detecting frame deletion, insertion and shuffling attacks through the use of image quality measure and a combination of temporal averages of non-overlapping sequence frames. This method has  an accuracy of 96\% and above for temporal attacks.  Similarly, other works address the temporal forgery in \cite{Gironi_2014, Jiang_2013}.

Additionally, another approach \cite{Ravi_2014} aims to use compression based techniques to identify forgeries within video frames. This was achieved through the extraction of compression noise then utilizing the transition probability matrices to classify whether a video had undergone forgery. Similarly, a novel approach was proposed in \cite{Su_2015} through the use of compressive sensing and K-SVD (k-Singular Value Decomposition) with an accuracy of 92.2\%. 

\paragraph{Sensor Noise Based Detection}
%utilizes fingerprint to detect forgeries

The use of sensor noise is utilized for both source identification as well as forgery detection methods. Accordingly, the works in \cite{Chetty_2010} utilizes a novel algorithm that extracts noise from both spatial and temporal features, utilizing them to detect forgeries with an accuracy of 92\%. A similar proposal was introduced in \cite{Hsu2_2008} as the authors extract noise residue through the use of a block-level correlation method to detect forgeries. Additionally, the authors in \cite{Hyun_2013} utilize sensor pattern noise and minimum average correlation energy filter to determine the authenticity of a video.

Additionally, another forgery detection algorithm is proposed in \cite{Kobayashi_2010}, it utilizes sensor noise to determine video forgeries through the use of pixel level analysis of the video's frames. Another work \cite{Pandey_2014} used sensor noise to determine whether a video had undergone copy-move forgery through the use of SIFT, as the sensor noise pattern would differ for an inserted object. Similarly, another work addresses copy-move forgery detection through sensor noise in \cite{Singh_2017}. Additionally, another proposed algorithm \cite{Singh2_2017} utilizes both sensor noise inconsistencies and pixel-correlation analysis to determine upscale-crop and splicing forgeries in forged videos.

\paragraph{Motion Features Based Detection}
%utilize time-dependent features in the video to determine the 

%\cite{Su_2015}

Motion features utilize time-dependent features in videos to determine if a forgery occurred. The approach proposed in \cite{Bestagini_2013} analyzes video sequences to detect the footprints left by forgeries, the authors utilized two methods, image-based and video-based, and compared consecutive frames. They were able to reach an accuracy of 92\%. Additionally, the authors in \cite{Bidokhti_2015} propose a passive technique to detect copy-move forgeries through the use of optical flow coefficient. Similarly, the work in \cite{Jia_2018} proposes to detect copy-move forgery based on optical flow.  Another approach \cite{Bozkurt_2017} detects forgeries by analyzing different frames and the correlations between them.

The approach proposed in \cite{Su_2009} focused on the detection of frame deletion through the use of motion-compensated edge artifact (MCEA). It is used to analyze the correlation between frames as detection method. Similarly, the authors in \cite{Dong_2012} propose the use of an MCEA method by analyzing changes in frames to detect video forgeries. Another work \cite{Chao_2012} proposes a novel inter-frame detection method that employs optical flow consistency to detect frame insertion and deletion. This method was able to detect forgeries and the forgery method. Another approach \cite{Feng_2014} proposes an algorithm that can determine both the existence and location of frame deletion through total motion residual analysis.

Another proposed algorithm in \cite{Karthikeyan_2020} used an automated method that combines MPEG-2 and optical flow in addition to block-based matching to detect forgeries. Another automated method is proposed by \cite{Kingra_2017} that utilizes optical flow and residual gradient providing a forgery detection method with 90\% accuracy. The authors in \cite{Tan_2015} propose an automated algorithm to identify object-based forgeries. Additionally, the works in \cite{Singh3_2017} propose the use of a hybrid algorithm that utilizes optical flow and prediction residual examination to detect temporal forgeries.

\paragraph{Statistical Features Based Detection}
%statistical and pixel, geometric and physics, inconsistencies (light)

Statistical based methods utilize attributes and pixels in video frames to determine forgeries. The authors in \cite{Lin_2012} examine the frame duplication forgery technique, they use candidate segment selection, spatial similarity measurement, classification and post-processing techniques. They achieved a DC of 95.1\%. The work in \cite{Althamneh_2016} proposes the use of local statistical information within the video to detect frame addition, removal and spatial tampering with a high accuracy in classification ranging from 90-100\%. 

Another approach is proposed in \cite{Aloraini_2019} that utilizes spatial decomposition, temporal filtering and sequential analysis to detect forgeries which are based on object changes. This method is resistant to compression and lower resolution videos. Additionally, the work in \cite{Aloraini_2020} proposes the use of both sequential and patch analysis to detect and localize the manipulated segments of a manipulated video. Additionally, the authors in \cite{Chitradevi_2014} proposed a forgery detection algorithm that utilizes mean frame comparison between groups of frames. Another approach in \cite{Jia_2015} used Automatic Color Equalization (ACE) to detect light changes to detect frame deletion.

The works in \cite{Lin_2014} used spatio-temporal coherence analysis to detect and localize forgeries on an object level. Additionally, the approach proposed in \cite{Mathai_2016} utilized statistical moment features and normalized cross correlation of the moment features to detect forgeries and localize them. Another work by \cite{Raskar_2021} proposes the use of atom structure analysis with a focus on MPEG-4 videos, with an accuracy of 80\% and above. Additionally, the authors in \cite{Su_2017} utilize exponential-fourier moments for detection of region duplication, with an accuracy of up to 93.1\%. Another work \cite{Su_2019} proposes the use of a novel method that utilizes energy factor and adaptive parameter-based visual background extractor algorithm. The proposed method shows an accuracy of up to 90.64\%.

In addition to these works, the authors in \cite{Tralic_2014} utilized cellular Automata and local binary patterns with block based method to detect copy-move forgeries. Another work in \cite{Ulutas_2018} utilized another algorithm, Bag-of-Words, to build a dictionary of visual words through the use of SIFT keypoints from the video to detect frame duplication forgeries. The approach proposed in \cite{Xu_2016} detects inter-frame forgeries by analyzing frames with the same background for frame continuity inconsistencies. Another work by \cite{Yin_2014} utilizes nonnegative tensor factorization to detect video forgeries, this algorithm is able to detect both frame deletion and insertion forgeries.

\paragraph{Machine Learning Based Detection}

Various works have proposed the use of machine learning and deep learning techniques to determine video forgeries. Accordingly, the algorithm proposed in \cite{Dvino_2017} utilizes auto-encoders and RNN (recurrent neural networks) to detect splicing forgeries. Additionally, another work \cite{Jaiswal_2013} utilizes SVM with a parameter identified through Prediction Error Sequence to detect temporal forgeries. Additionally, another work in \cite{Shanableh_2013} utilizes machine learning to detect frame deletion through video bit stream and image features. 

The work in \cite{Johnston_2020} utilized a deep learning algorithm on authentic videos to learn the features of authentic videos and determine their key frames, then the algorithm is used on tampered videos to localize tampered regions. Another approach proposed by \cite{Saddique_2019} used irregularities of video texture and micro-patterns in video frames with SVM and consecutive frame analysis to determine spatial forgeries. This method has a detection accuracy of up to 96.68\%. Additionally, the work in \cite{Zampoglou_2019} proposed an automated forgery detection algorithm that utilized DCT and CNN.

\subsection{Generative Video Forensics}\label{video_gen}
%% Face Synthesize
%% Identity Swap
%% Attribute Manipulation
%% Expression Swap
%survey, CGI 
%\cite{Tolosana2_2020, Yao2_2018}

%image
%\cite{Chengi_2021}

%GAN
%\cite{Clark_2019, Elkhiyari_2016}

Due to the increased use of and accessibility to applications that enable the creation of generative videos. Synthesized or generative videos undergo different types of forgeries. These forgeries include the below \cite{Tolosana_2020}:
\begin{itemize}
  \item Face Synthesize: This technique entails the creation of a fake face through the use of generative models.
  \item Identity Swap: This method swaps the face of a person for another person's face.
  \item Attribute Manipulation: This tampering technique alters the face in the video with retouching effects, e.g. smoothing skin or changing hair color.
  \item Expression Swap: This method changes the expression of the person in the video.
\end{itemize}

Several works examine generative forgery detection as shown in Table \ref{GVideoForgery}, the work in \cite{Amerini_2019} introduces the use of optical flow field and CNN to identify inter-frame dissimilarities to classify Deep Fakes or face swapping techniques. Another work in \cite{Demir_2021} proposes the use of biological markers to identify tampered videos with a focus on eye and gaze features. This approach was able to detect Deep Fakes with 80\% in the wild. Similarly, a deep learning approach is proposed in \cite{Jung_2020} to detect deep fakes through the use of human eye blinking patterns with an accuracy of 87.5\%. A similar method was presented in \cite{Li5_2018}. Another work in \cite{Guera_2018} utilizes CNN to extract frame-level features in addition to the use of RNN to classify if a video had been tampered. 

In addition, another method is introduced in \cite{Koopman_2018} that analyzes PRNU in generative videos and has shown that the patterns of real video and manipulated videos vary. This shows that PRNU may potentially be used to identify generative videos. Another algorithm is proposed in \cite{Korshunov_2018} that showed that face swapping GANs are challenging to be recognized by both face recognition and audio-visual tools. Similarly, another work \cite{Li3_2018} utilizes 3D CNN to identify face manipulation and spoofing. Another method presented in \cite{Yang_2019} used inconsistent head pose cues and SVM to detect face forgeries. 

Furthermore, another work in \cite{Kumar_2020} utilizes deep learning based on metric learning to classify deep fakes that have high compression factors, with an accuracy of at least 90.91\%. Another approach in \cite{Li4_2018} proposes the use of CNN to detect deep fakes through the distinct features remaining on videos due to the limited resolution of Deep Fakes, it had been tested on various data sets with an accuracy of least 93.2\%. A similar work was presented in \cite{Nirkin_2021}. Another work \cite{Qian_2020} utilizes collaborative learning for deep neural networks and mining forgery patterns to detect face forgeries in videos. Additionally, the work in \cite{Bonomi_2021} utilized dynamic texture analysis to distinguish face forgeries through the use of local derivative patterns on three orthogonal places and SVM classifier.

Additionally, an approach was proposed in \cite{Cozzolino_2021} that aims to identify facial manipulations in a generalized manner. The method trained two neural networks on a data set that does not contain manipulated videos, as such any other manipulated video would be classified by the algorithm. Similarly, the works in \cite{Luo_2021} aims to identify multiple generative video attacks through a white box method which includes multiple adversarial samples within the Deep Learning model. The authors in \cite{Sabir_2019} proposed the use of RNN to discern temporal information in a video and detect forgeries including Deep Fakes, face2face and face swapping methods, with an accuracy of at least 94.35\%. Additionally, the work in \cite{Zheng_2021} utilizes a fully temporal convolution network to utilize temporal coherence for a more generalized face forgery detection technique.

\begin{table}[]
\begin{tabular}{|p{25mm}|p{45mm}|p{30mm}|p{30mm}|}
\hline
Forgery Detection Technique & Algorithms & Pros & Cons   \\ \hline
Machine Learning & Physical Markers \cite{Yang_2019}, Dynamic Texture \cite{Bonomi_2021} & It can be used to detect generative forgeries. & It has not been shown to have generalized applications.   \\ \hline
Deep Learning & Optical Flow \cite{Amerini_2019}, Biological Markers \cite{Demir_2021, Jung_2020, Li5_2018}, Temporal Frame \cite{Guera_2018, Sabir_2019, Zheng_2021}, Mining \cite{Qian_2020}, Compression Based \cite{Kumar_2020, Li4_2018, Nirkin_2021} Generalized \cite{Li3_2018}, Sensor Noise \cite{Koopman_2018, Cozzolino_2021, Luo_2021} & It can be used to detect various forgeries and has been used for generalized forgery detection. & Deep learning methods are susceptible to AF.  \\ \hline
\end{tabular}
\caption{Overview of Generative Forgery Detection Techniques}
\label{GVideoForgery}
\end{table}

\subsection{Video Anti-Forensics}\label{Video_anti-forensics}

%\cite{Ding_2021, Stamm2_2011, Stamm_2012, Stamm2_2012}
The aim of video anti-forensics is to challenge the forgery detectors and conceal video manipulations. The works in \cite{Stamm2_2011} proposes an anti-forensics technique to conceal frame deletion by removing temporal fingerprint artifact in MPEG videos. A similar method is proposed in \cite{Stamm2_2012}, which focused on temporal forensics in both frame deletion and addition.  Another approach proposed in \cite{Ding_2021} used an anti-forensics algorithm that utilizes a GAN model that consists of supervising modules that aim to enhance the quality of Deep Fakes without noticeable artifacts. This method was shown to deceive forensics detectors. Additionally, the work in \cite{Liu_2022} proposes the use of a more generalized anti-forensics method that can deter multiple forensics detectors that detect trace removal attacks. The method focuses on navigating spatial anomalies, spectral anomalies and noise fingerprint methods of forensic detection through the introduction of trace removal network during the creation of the generated image.

\subsection{Video Counter Anti-Forensics}\label{video_counter}
%\cite{Kang_2016}
The counter anti-forensics field aims to mitigate the anti-forensics techniques and detect the use of these methods. Accordingly, the work in \cite{Kang_2016} proposes a counter anti-forensics approach for inter-frame forgeries, by analyzing the anti-forensics temporal forgery method. The authors were able to detect remaining artifacts that can be used to detect video forgeries. This method can be used for both frame insertion and deletion anti-forensics techniques.

\section{Open Problems and Conclusion} \label{Problems_Conclusion}
%- The lack of AF and CAF in video forensics
%- The advancement and difficulties in identifying GANs
%- The difficulty to identify fake media
%- How OSN hide important artifacts when processed
%- Needing combination of general purpose and signature based solutions
%- Admissibility in court (real life application)

This paper explores both image and video forensics which includes source identification and forgery detection for camera generated images and videos. It also explores generative images and videos which are starting to be indistinguishable from real photos. The paper also delves into anti-forensics and counter anti-forensics for both image and video forensics. However, it does not discuss audio and text forensics, which is a potential area for research. Additionally, the literature demonstrates that there are a number of open problems in the image and video forensics field.

\paragraph{Challenges due to Anti-Forensics.} It is important to indicate that many of the techniques used for image authenticity and manipulation detection are weak to adversarial attacks, which shows that they are susceptible to anti-forensic attacks. Hence, the upcoming forgery detectors must take into consideration the placement of counter anti-forensics. Moreover, the surveyed anti-forensics and counter anti-forensics techniques appear to be very specific as they aim to circumvent certain techniques and the counter anti-forensics mainly aim to counter each proposed anti-forensic technique. Therefore, further research can explore more general-purpose methods to counter the anti-forensic techniques which include proposing forensic detectors that have counter anti-forensics built within to circumvent adversarial attacks.

\paragraph{Generalized Forgery Detection Techniques.} It has been shown that forgery detection methods mainly rely on signature-based inference. Recently, more novel approaches are being introduced in the field to detect unknown manipulations. These approaches can be used to detect various manipulations and appear to have the most potential for future research. 

\paragraph{Challenges in GAN detection.}Furthermore, there appears to be a lack of research to discern whether an image is real or forged or is the result of a GAN. The field of image generative images detection is limited and that may be due to most of the research efforts related to generative images detection being focused on video forensics. Future digital image forensics detectors will need to include generative image detection as the use of this technology is increasing in image creation. 

\paragraph{Lack of Video Anti-Forensics and Counter Anti-Forensics.} The literature shows that the video forensics field lacks literature on video anti-forensics and counter anti-forensics as most anti-forensics and counter anti-forensics efforts and literature are focused on images.

\paragraph{Legalities and Admissibility of Detection Techniques.} Additionally, the field of forensics aims to provide evidence of the authenticity of multimedia to ensure that the evidence is admissible in a court of law; therefore, forensic experts must be able to justify the authenticity of the evidence. However, with the research direction heading toward deep learning, it is difficult for forensic experts to discern the reason why an image was classified as authentic or fake, which may pose an issue for experts in the forensic field \cite{Meena_2019, Bohme_2013, Yang_2020, Marra2_2018, Mandelli_2020, Fernando_2021, Dehnie_2006}.

\printbibliography
\end{document}